\documentclass[twocolumn,showpacs,preprintnumbers,superscriptaddress,amsmath,amssymb]{revtex4-1}

\usepackage[colorlinks=true,bookmarks=false,citecolor=blue,urlcolor=blue]{hyperref} 

\usepackage{amsmath,amssymb}
\usepackage{graphicx}
\usepackage{verbatim}
\usepackage{enumitem}
\usepackage{color}
\usepackage[english]{babel}
\usepackage{bm}
\usepackage{natbib}
\def\dfrac{\displaystyle\frac} 
\newcommand{\eps}{\varepsilon}
\usepackage{epstopdf}
\usepackage{setspace}
\usepackage{comment}

\definecolor{lavenderpurple}{rgb}{0.59, 0.48, 0.71}

\begin{document}

\title{Optical anapoles in nanophotonics and meta-optics}

\author{Kseniia V. Baryshnikova}
\affiliation{ITMO University, St.~Petersburg 197101, Russia}
\author{Daria A. Smirnova}
\affiliation{Nonlinear Physics Center, Australian National University, Canberra ACT 2601, Australia}
\affiliation{Institute of Applied Physics, Russian Academy of Sciences, Nizhny Novgorod 603950, Russia}
\author{Boris S. Luk'yanchuk}
\affiliation{Faculty of Physics, Lomonosov Moscow State University, Moscow 119991, Russia}
\affiliation{School of Physical and Mathematical Sciences, Nanyang Technological University, 637371 Singapore}
\author{Yuri S. Kivshar}
\affiliation{ITMO University, St.~Petersburg 197101, Russia}
\affiliation{Nonlinear Physics Center, Australian National University, Canberra ACT 2601, Australia}



\begin{abstract}
Interference of electromagnetic modes supported by subwavelength photonic structures is one of the key concepts that underpins the subwavelength control of light in meta-optics.  It drives the whole realm of all-dielectric Mie-resonant nanophotonics with many applications for low-loss nanoscale optical antennas, metasurfaces, and metadevices. Specifically, interference of the electric and toroidal dipole moments results in a very peculiar, low-radiating optical state associated with the concept of {\em optical anapole}. Here, we uncover the physics of multimode interferences and multipolar interplay in nanostructures with an intriguing example of the optical anapole. We review the recently emerged field of {\it anapole electrodynamics} 
explicating its relevance to multipolar nanophotonics, including direct experimental observations, manifestations in nonlinear optics, and rapidly expanding applications in nanoantennas, active photonics, and metamaterials. 
\end{abstract}

\maketitle

\section{Introduction}

During last decade, we observe a growing interest to the study of various optical effects associated with the so-called {\em toroidal electrodynamics}~\cite{Zheludev2016}, in particular, in the field of metamaterials and nanophotonics. This interest largely originates from many opportunities provided by multipolar response  of subwavelength optical structures, allowing a deeper insight into many optical phenomena at the nanoscale, including multipolar nonlinear nanophotonics~\cite{Smirnova2016}. 
Conventional multipole expansions~\cite{BohrenHuffman,Jackson1965,Grahn2012}, applied to dynamic charge–current distributions on length scales  much smaller than the effective wavelength of light,  feature dominant cartesian electric and magnetic dipoles. Nevertheless, when the dynamic charge–current is distributed within an area comparable to or larger than the effective wavelength of light, higher-order multipoles including dynamic toroidal multipoles become non-negligible and contribute to optical response.

The physics of interference and interplay between three families of  multipolar modes~\cite{Radescu2002}: electric, magnetic and toroidal ones, underpins nanoscale light manipulation. Indeed, toroidal multipoles provide physically significant contributions to the basic characteristics of matter including absorption, dispersion, and optical activity~\cite{Zheludev2016}. In the far field zone, they produce the same radiation patterns as corresponding electric or magnetic multipoles.  It can lead to destructive interference effects and vanishing scattering accompanied with highly nontrivial field distribution and strong light confinement.  Perhaps, the simplest example of this kind is the interference of the electric dipole (ED) and toroidal dipole (TD) modes.  The toroidal dipole corresponds to currents flowing on the surface of a torus. Since both EDs and TDs have identical radiation patterns, when co-excited and spatially overlapped with the same radiation magnitude but out of phase, they cancel the scattering of each other in the far field region, and the scatterer appears to be invisible. Such a radiationless state with nontrivial oscillating current configuration (in analogy with non-oscillating poloidal current) was termed {\em anapole} (from Greek ``ana'', ``without'', thus meaning ``without poles''). 
In fact, this term, as well as toroidal moments, was first theoretically introduced by B.~Zeldovich back in 1957~\cite{Zeldovich1957}, and  also occurs in the nuclear~\cite{Wood1997} and dark-matter physics~\cite{HoPLB2013,Gao2014darkmatter}. In the context of classical electrodynamics, an anapole mode corresponds to a specific type of the charge-current distribution that neither radiates nor interacts with external fields. In a broader context, in the scattering theory the anapole is referred to suppression of scattering independently of the multipoles involved.

In subwavelength dielectric particles,
the optical response is governed by the induced polarization currents which generate different multipoles of the electromagnetic field and are capable of producing {\em optical anapole modes}.  Such weakly radiating modes have great potential for numerous applications in nanophotonics and meta-optics, which motivates this review. We discuss the basic properties of these special optical states, and review briefly the recent progress in the field of {\em anapole electrodynamics} with applications to nanophotonics, nonlinear optics, and metamaterials.

The concept of anapole represents a specific example of a plethora of interference effects which can occur in subwavelength dielectric particles. During recent years, several related concepts have emerged, and they show their importance for applications of all-dielectric Mie-resonant photonics. We discuss some of those effects in the concluding section of this review paper. 

\section{Anapole concept} 

The basic physics of optical anapoles can be qualitatively explained by the Mie theory describing the light scattering by spherical particles~\cite{BohrenHuffman,Jackson1965}. Within this theory, the scattering efficiency, $Q_{\text{sca}}$, extinction, $Q_{\text{ext}}$, and other characteristics of the optical response are expressed through the electric, $a_l$, and magnetic, $b_l$, scattering amplitudes
\begin{subequations}
\label{eq:Qsca}
\begin{align}
Q_{\text{sca}}=\frac{2}{q^2}\sum\limits_{l=1}^\infty (2l+1)\left(|a_l|^2+|b_l|^2 \right)\:,\\
Q_{\text{ext}}=\frac{2}{q^2}\sum\limits_{l=1}^\infty (2l+1)Re\left(a_l+b_l\right) \:.
\end{align}
\end{subequations}
Here, the efficiencies are defined as ratios of the corresponding cross-section to the geometrical cross-section $\sigma_{\text{geom}}=\pi R^2$, where $R$ is radius of the particle. The symbol $q$ denotes the so-called size parameter defined as $q=2\pi R/\lambda$, and $l$ is the index numbering the orbital modes: dipolar $l=1$, quadrupolar $l=2$, octupolar $l=3$, etc. The scattering amplitudes, corresponding to the partial spherical waves, are given by  
\begin{subequations}
\label{eq:Qsca}
\begin{align}
a_l=\dfrac{F^{(a)}_l}{F^{(a)}_l + i G^{(a)}_l }\:,\\
b_l=\dfrac{F^{(b)}_l}{F^{(b)}_l + i G^{(b)}_l }\:,
\end{align}
\end{subequations}
where for the spherical non-magnetic particle of the refractive index $n\equiv\sqrt{\eps}$ in vacuum, the quantities ${F^{(a,b)}_l}$ and ${G^{(a,b)}_l}$ are defined as
\begin{equation}
\begin{aligned}
&{F^{(a)}_l}=n \psi^{'}_l(q) \psi_l (nq) - \psi_l(q) \psi^{'}_l(nq)\:,\\
&{G^{(a)}_l}=n \chi^{'}_l(q) \psi_l (nq) - \psi^{'}_l(nq) \chi_l(q)\:,\\
&{F^{(b)}_l}=n \psi^{'}_l(nq) \psi_l (q) - \psi_l(nq) \psi^{'}_l(q)\:,\\
&{G^{(b)}_l}=n \chi_l(q) \psi^{'}_l (nq) - \psi_l(nq) \chi^{'}_l(q)\:.
\end{aligned}
\end{equation}
Here, the functions 
\[ \psi_l(q) = \sqrt{\dfrac{\pi q}{2}} J_{l+\frac{1}{2}} (q), \;\;\;\;  \chi_l(q) = \sqrt{\dfrac{\pi q}{2}} N_{l+\frac{1}{2}} (q) \]
are expressed through the Bessel and Neumann functions~ \cite{BohrenHuffman}. 

\begin{figure}[t!] 
\centering\includegraphics[width=0.99\linewidth]{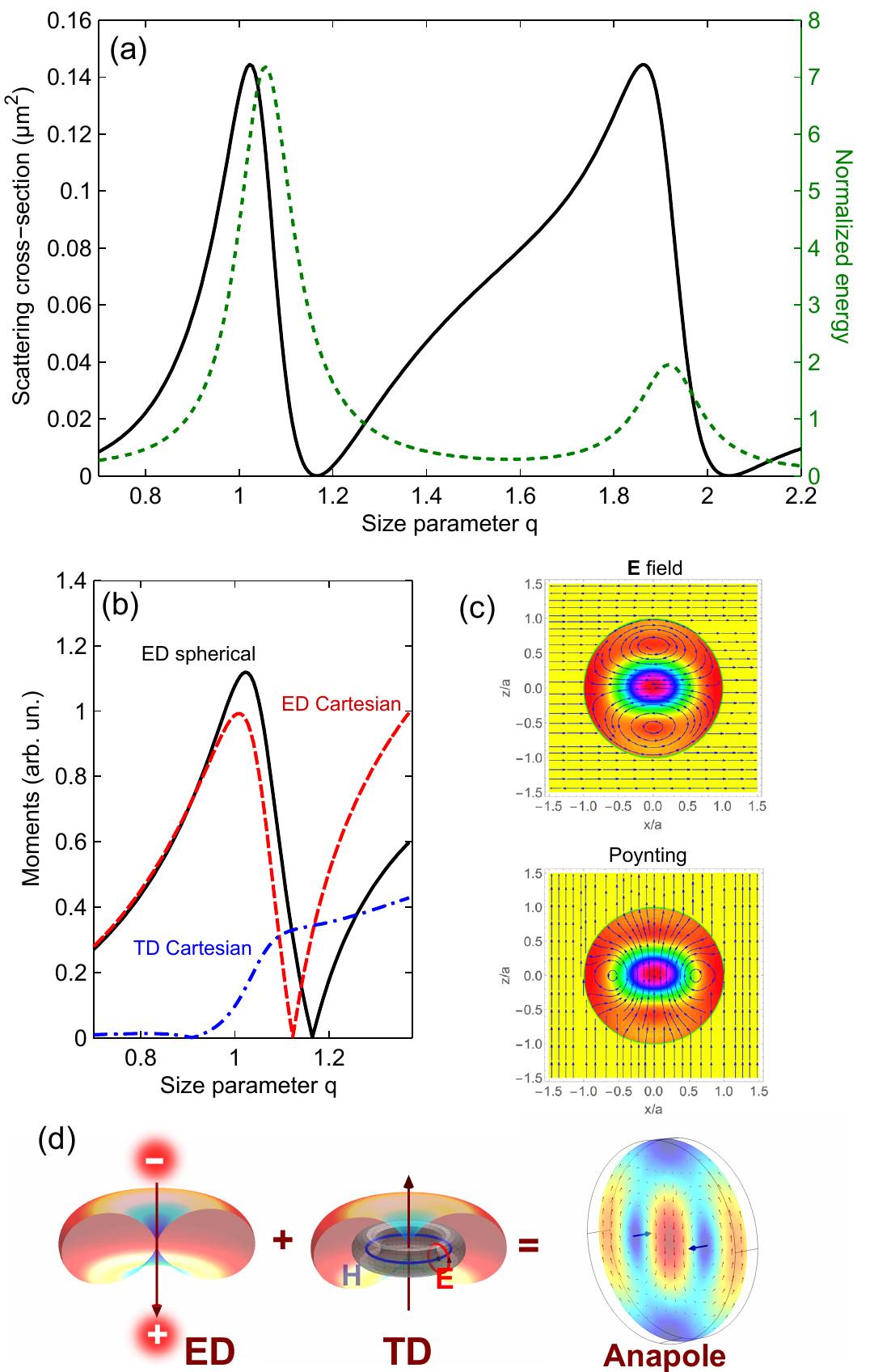} 
 \caption{(a) 
Scattering cross-section (black) and normalized time-averaged energy density (dashed green) of the electric dipole field excited inside a dielectric spherical particle of 
refractive index $n=4$ at wavelength $550$~nm. 
(b) Calculated spherical electric dipole $|{\bf p}|$ (black), Cartesian electric $|{\bf p}_{\text{Car}}|$ (red) and toroidal  $|-ik {\bf T}_{\text{Car}}|$ (blue) dipole moments  for a dielectric particle of refractive index $n=4$ in vacuum at wavelength $550$~nm.
(c) The poloidal current distribution of the vector electric field (top) inside the particle and distribution
of the Poynting vector (bottom) at $q=1.1654$.
(d) Anapole concept: the destructive interference of the electric (ED) and toroidal (TD) dipole moments with identical doughnut-shaped radiation patterns and differing charge-current distributions suppresses the far-field scattering 
to create an  optical anapole mode in the silicon nanodisk (calculated numerically for parameters of Ref.~\cite{miroshnichenko2015nonradiating}). The electric dipole moment is illustrated as a separation of positive and negative electrical charges. The toroidal dipole moment is associated with with the circulating magnetic field ${\bf H}$ and poloidal electric field $\bf E$ distribution. The mode profile is shown at midheight cut of the disk.  
}
\label{fig:concept}
\end{figure}

First, we discuss the zeros of the electric dipole scattering defined by the condition, $a_1=0$. This condition corresponds to 
$ F^{(a)}_l = 0$, and it is fulfilled along the trajectory~\cite{luk2017hybrid}, where we should consider only those solutions with $\cos(q) \ne 0$ and $\cos(nq) \ne 0$.
For non-magnetic materials, this condition corresponds to the following equation,
\begin{multline}
1-n^2+q(n^2-1+n^2q^2)cot(q)+nq(n^2-1-n^2q^2)cot(nq)+\\nq^2(1-n^2)cot(q)cot(nq)=0\:.
\end{multline}
For example, at $n=4$ the first zero is reached at $q=1.1654$, and, as explained below, arises due to the destructive interference of the Cartesian electric and toroidal dipoles radiation~\cite{miroshnichenko2015nonradiating,luk2017hybrid}. As an example, in Fig.~\ref{fig:concept}(a-c) we show this “ideal anapole” which originates from the Mie theory, where the only electric dipole contribution is taken into account in the scattered field. One can easily identify the poloidal current distribution of the electric vector inside the nanoparticle in Fig.~\ref{fig:concept}(c). However, under influence of (always present) higher-order modes, this picture becomes less ideal. This problem was discussed in Ref.~\cite{luk2017hybrid} in detail. We notice that there is almost no energy flow in the region inside the loops of the separatrices around the singular points of the Poynting vector distribution. Such an electric anapole can be considered as being equivalent to a confocal system of two lenses with the numerical aperture (NA) close to unity~\cite{luk2017hybrid}. 

\begin{table*}[t]
  \centering
  \caption{{\bf Expressions for exact spherical and Cartesian dipole moments.}~\cite{alaee2018electromagnetic,Li2018,wu2018optical}}
  \label{tab:table2}
  \begin{tabular}{l|c|c}
    l=1  & spherical &  Cartesian \\
    \hline
    {ED} & \parbox[t]{9.5cm} { $\mathbf{p}=-\dfrac{1}{i\omega}\Bigl\{\int \mathbf{J}  j_0(kr)  d^3 \mathbf{r}  + \dfrac{k^2}{2}\int \left(3(\mathbf{r}\cdot\mathbf{J})\mathbf{r} -  r^2  \mathbf{J}\right)\frac{j_2(kr)}{(kr)^2} d^3\mathbf{r}\Bigr\}$ \\ $\approx \mathbf{p}_{\text{Car}} + ik \mathbf{T}_{\text{Car}} + \dfrac{k^2}{10}\mathbf{T}^{(1)} + ...$}  & \parbox[t]{6cm} {$\mathbf{p}_{\text{Car}}=-\dfrac{1}{i\omega} \int \mathbf{J} d^3 \mathbf{r} $ \\ $\mathbf{T}_{\text{Car}}=\dfrac{1}{10c}\int \left((\mathbf{r}\cdot\mathbf{J})\mathbf{r} - 2 r^2  \mathbf{J}\right) d^3\mathbf{r}$ \\ $\mathbf{T}^{(1)} = \dfrac{1}{28c}\int r^2 \left( 3 r^2 \mathbf{J} - 2 (\mathbf{r}\cdot\mathbf{J})\mathbf{r}   \right) d^3\mathbf{r}$}  \\ 
      \hline
          MD &  \parbox[t]{5cm}{$\mathbf{m}=\dfrac{3}{2}\int \left[\mathbf{r}\times\mathbf{J} \right] \dfrac{j_1(kr)}{kr} d^3\mathbf{r}$ \\ 
          $\approx \mathbf{m}_{\text{Car}}- \dfrac{k^2}{10}\mathbf{m}^{(1)} + ...$}  &  \parbox[t]{5cm} {$\mathbf{m}_{\text{Car}}=\dfrac{1}{2}\int \left[\mathbf{r}\times\mathbf{J} \right] d^3\mathbf{r}$ \\ $ 
      \mathbf{m}^{(1)}    =\dfrac{1}{2c}\int r^2\left[\mathbf{r}\times\mathbf{J} \right] d^3\mathbf{r}$ } 
  \end{tabular}
\end{table*}

In a similar way, one can find the conditions for the magnetic dipole amplitude satisfying the relation $b_1 = 0$ (the so-called {\it magnetic anapole}). However, in contrast to the electric anapole for which the energy is trapped inside the subwalength particle, in the case of the magnetic anapole the energy is expelled outside the particle with the near-field enhancement similar to that usually associated with plasmonic particles~\cite{luk2017hybrid}. Both types of anapole modes can be excited simultaneously, resulting in the formation of {\it hybrid anapole modes}. For spherical particles, this effect is hindered by the partial contributions from higher-order modes, which lead to the formation of nontrivial field configurations with vortex-like structures and singularities inside and outside the particle. 
Nevertheless, we expect that this restriction could be removed for other geometrical shapes (such as disks or spheroids) or multilayered particles.  Using the same approach, the anapole concept, associated with nontrivial nonradiating current distributions, can further be generalized to any higher order expansion of the Mie theory (e.g., for vanishing $a_2$ coefficient, we should realize {\it an electric quadrupolar anapole} when the Cartesian electric quadrupole mode compensates the toroidal quadrupole mode~\cite{Butakov2016,wu2018optical}). 

In the Rayleigh limit, any subwavelength particle effectively scatters light as an electric dipole, and the scattering increases with the fourth power of the frequency according to the well-known formula~\cite{Born},
\begin{equation}
Q_{\text{sca}}^{\text{(Ra)}}=\frac{8}{3}\left |\frac{\varepsilon-1}{\varepsilon+2}\right |^2q^4\:.
\end{equation}
It implies that electric dipole scattering is always present. However, by employing subwavelength particles with high values of the refractive index, it is possible to suppress the scattering below the Rayleigh approximation with the anapole mode excitation~\cite{Lukyanchuk_suppression_2017}. 
The scattering pattern near the anapole condition is not rotationally symmetric, thus yielding polarization-dependent scattering (except, of course, in the backward and forward directions)~\cite{Lukyanchuk_suppression_2017,savinov2018light}. This behavior is quite similar to the change of directivity in the vicinity of quadrupole resonance within weakly dissipating plasmonic particles~\cite{Lukyanchuk_2010}. While the spherical geometry allows suppressing the electric and magnetic dipole moments, it does not allow doing the same with the quadrupole and octupole modes simultaneously. Nevertheless, by using nonspherical geometries or layered structures, it could be possible to simultaneously suppress higher-order multipolar contributions. For example, further minimization of the scattering cross section is possible if one considers spheroidal~\cite{Lukyanchuk_2015} or disk-shaped particles~\cite{miroshnichenko2015nonradiating}. 
Practical applications of optical anapole structures are mostly  related to non-spherical scatterers. Design of such structures needs the multipole expansion~\cite{Jackson1965} of the electromagnetic fields of a sub-wavelength system of charges and currents.

\vspace{-0.1 cm}

Multipolar decomposition can be carried out in both spherical and Cartesian bases~\cite{Evlyukhin2016,Terekhov2017,zenin2017direct,alaee2018electromagnetic,Li2018} based on light-induced polarization or displacement current, $\mathbf{J} (\mathbf{r}) = - i \omega \mathbf{P}  (\mathbf{r}) = - i \omega \eps_0 (\eps - 1) \mathbf{E} (\mathbf{r})$, where $\mathbf{E} (\mathbf{r})$ is the total electric field inside the particle. 
In the context of a multipole expansion, the origin of toroidal multipole moments can be understood from first principles. Exact multipolar expansion in spherical basis does not use, contrary to the Taylor expansion in Cartesian coordinates, the approximation of small sources in the long-wave length limit. Over this approximation the derivation procedure becomes more intricate, that may be a reason why toroidal multipoles were overlooked often in Cartesian Taylor expansion. With the canonical spherical basis, one can get the three multipole moment families in a more direct and natural way~\cite{Nanz2016,alaee2018electromagnetic}. The radiated field (up to the quadrupolar order $l=2$) is related to the induced multipole moments as  
\begin{multline}
\mathbf{E}_{\text{sca}}(\mathbf{n})\approx \frac{k_0^2}{4\pi\varepsilon_0} \dfrac{e^{ik_0r}}{r} \left([\mathbf{n}\times[\mathbf{p}\times\mathbf{n}]]+\frac{1}{c}[\mathbf{m}\times\mathbf{n}]\right.\\-\left.\frac{ik_0}{6}[\mathbf{n}\times[ \mathbf{Q}^{e}\times \mathbf{n} ]]-\frac{ik_0}{2c}[ \mathbf{Q}^{m} \times \mathbf{n}] + ... \right) \:,
\end{multline}
where $\mathbf{p}$, $\mathbf{m}$, $\mathbf{Q}^{e}$, $\mathbf{Q}^{m}$ are the spherical multipole moments expressed in the Cartesian coordinates. The exact expressions for these moments valid for any particle’s size and shape are derived in Ref.~\cite{alaee2018electromagnetic}. 
The approximate Cartesian multipole moments are obtained in the long-wavelength approximation  
by expanding each multipole term with respect to $kr$ in the Bessel functions entering the exact expressions, and correspond to the leading order.  Thereby, the toroidal multipoles are retrieved as the next-order terms in these expansions~\cite{alaee2018electromagnetic,Li2018,wu2018optical}. In particular, we get ${\mathbf p} \approx \mathbf{p}_{\text{Car}} + i k \mathbf{T}_{\text{Car}}$, where
Cartesian electric dipole $\mathbf{p}_{\text{Car}}$ and toroidal dipole $\mathbf{T}_{\text{Car}}$ moments are given in Table~\ref{tab:table2}, where  
$j_{1,2,3}$ is spherical Bessel function of the first, second, and third order, respectively, 
and integration is performed over the nanoparticle volume.  
Neglecting higher-order terms, including  mean-square radii multipoles $\mathbf{m}^{(1)}$ and $\mathbf{T}^{(1)}$~\cite{dubovik1990toroid,Li2018}, 
the 
scattering cross-section in vacuum can be then approximately calculated using  Cartesian moments as \cite{Evlyukhin2016} 
\begin{multline}
\sigma_{\text{sca}}\approx \frac{k_0^4}{6\pi\varepsilon_0^2|\mathbf{E}_{\text{inc}}|^2}\left|\mathbf{p}_{\text{Car}}+
ik\mathbf{T}_{\text{Car}}\right |^2+
\frac{k_0^4\mu _0}{6\pi \varepsilon _0 |\mathbf{E}_{\text{inc}}|^2}|\mathbf{m}_{\text{Car}}|^2\\+
\frac{k_0^6}{720\pi\varepsilon_0^2|\mathbf{E}_{\text{inc}}|^2}\sum\limits|\hat{Q}^{e}_{\text{Car}}|^2+
\frac{k_0^6\mu_0}{80\pi\varepsilon_0|\mathbf{E}_{\text{inc}}|^2}\sum\limits|\hat{Q}^{m}_{\text{Car}}|^2 + ... 
\end{multline}

The fundamental character of toroidal moments seems to be
questionable in electrodynamics, as they mathematically appear as terms in series expansions. However, toroidal moments are a useful concept for describing certain charge-current distributions. Remarkably, three families of multipoles are complementary because they represent distinct physical symmetry
properties~\cite{Nanz2016}. For instance, for dipole moments: $\bf{p}_{\text{Car}}$ is even under time reversal, but odd under spatial inversion, $\bf{m}_{\text{Car}}$  is odd under time reversal, but even under spatial inversion, and $\bf{T}_{\text{Car}}$ is odd under both time and spatial inversion. 

Thus, the first zero of the electric dipole scattering by a spherical particle, for example, can be explained in terms of destructive interference of Cartesian electric and toroidal dipoles, as shown in Fig.~\ref{fig:concept}. 
For small particles, 
the spherical and Cartesian electric dipoles are almost identical and the toroidal moment is negligible, whereas for larger sizes, the contribution of the toroidal dipole moment
has to be taken into account [see Fig.~\ref{fig:concept}(b)]. The anapole state 
corresponds to the vanishing spherical electric dipole, where the Cartesian electric and toroidal dipoles cancel each other, $\mathbf{p}_{\text{Car}} = - i k \mathbf{T}_{\text{Car}}$. At the same time, the electromagnetic energy inside the particle is not zero, implying that there exists some non-trivial excitation, see Fig.~\ref{fig:concept}.  A similar example of the fundamental anapole state composed of a toroidal dipole moment and electric dipole moment excited in the dielectric nanodisk is shown in Fig.~\ref{fig:concept}(c). 
The series of zeroes in electric dipole scattering coefficient can be treated as anapoles of the increasing order, but for higher-order anapoles additional terms beyond the toroidal dipoles  in the expansions of multipolar moments are required~\cite{alaee2018electromagnetic,Li2018,zenin2017direct}.

As an alternative to the toroidal approach and multipolar decomposition, the formation of anapole states in pure
dielectric nanoparticles can be described in terms of the resonant-state expansion. In this way, the anapole behavior can be interpreted as a result of a Fano-type interference between different resonant modes with complex eigenfrequencies~\cite{Stout2017,Powell2017}. The excitation of an anapole mode is associated with the characteristic Fano lineshape in the spectrum~\cite{Lukyanchuk_suppression_2017}. 
The high-order radiation-less states can be also deduced as the result of a complex interaction among the resonances (modes) of the cavity and the surrounding environment, which are mutually coupled by boundary terms, by applying a Fano-Feshbach projection scheme~\cite{gongora2017fundamental}, a theoretical method widely used in quantum mechanics of open systems.

\section{Experimental observations}

The anapole mode in optics can be viewed as an engineered
superposition of toroidal and electric or magnetic multipoles,
resulting in destructive interference of the radiation fields.
The toroidal modes involved into interference can be generated in specifically designed composite metamaterials~\cite{Kaelberer2010,Fedotov2013,Zheludev2016,sayanskiy2018anapole,cong2018metamaterial,talebi2018theory,guo2014electric,li2014resonant,li2015low,liu2015toroidal},  where the near-field coupling between the individual particles' modes is capable of suppressing all standard multipoles with the electromagnetic scattering due to the resulting 
toroidal excitations being a dominant physical mechanism of the metamaterial response.
Isolated excitation of toroidal multipoles requires careful
engineering of the structure to match the
special corresponding near-field current distributions.
Particular designs were proposed and fabricated that experimentally verified the anapole
states in microwave metamaterials~\cite{basharin2017extremely} and metamolecules~\cite{nemkov2017nontrivial}, as well as at at optical frequencies~\cite{wu2018optical}.
Still, even simple homogeneous dielectric particles with
Mie resonances can support toroidal dipole moments,
co-exciting with the electric and magnetic multipoles.  

To observe a pure optical anapole, e.g. based on ED and TD, besides the proper excitation of ED and TD modes, the suppression of other
multipoles is required. 
In the discussion above we considered a hypothetical situation of single partial spherical wave excitation, which is quite difficult to realize in practice. 
It is not possible to detect the anapole mode in homogeneous spheres with
incident plane waves since the MDs and other quardupolar excitations at the ED–TD scattering
cancellation point are non-eligible.
However, under proper excitation, the anapole state can be 
generated even in simple isolated scatterers, such as dielectric spheres, disks, nanorods~\cite{Liu2015}. 
Specifically, the pure anapole excitation can be achieved
with other unwanted multipoles suppressed within the following individual nanoparticles in the
optical regime: (i) all-dielectric nanodisks with incident plane waves~\cite{miroshnichenko2015nonradiating};
(ii) all-dielectric spheres with engineered incident waves (two counter propagating radially
polarized beams with the same intensity but out of phase)~\cite{wei2016excitation}; 
and (iii) core–shell metal–dielectric nanospheres~\cite{liu2015toroidal} or nanorods~\cite{LiuOL2015} and homogeneous spherical particles possessing the radial
anisotropy~\cite{LiuJN2015,LiuOE2015} with incident plane waves . 
Excitation of anapole modes in silicon nanoparticles can be facilitated by using structured light~\cite{Lamprianidis2018}. Particularly, azimuthally polarized focused beams can be employed to achieve the magnetic anapole modes corresponding to the suppression of magnetic dipole scattering. 

\begin{figure*}[!t]
\centering \includegraphics[width=0.9\linewidth]{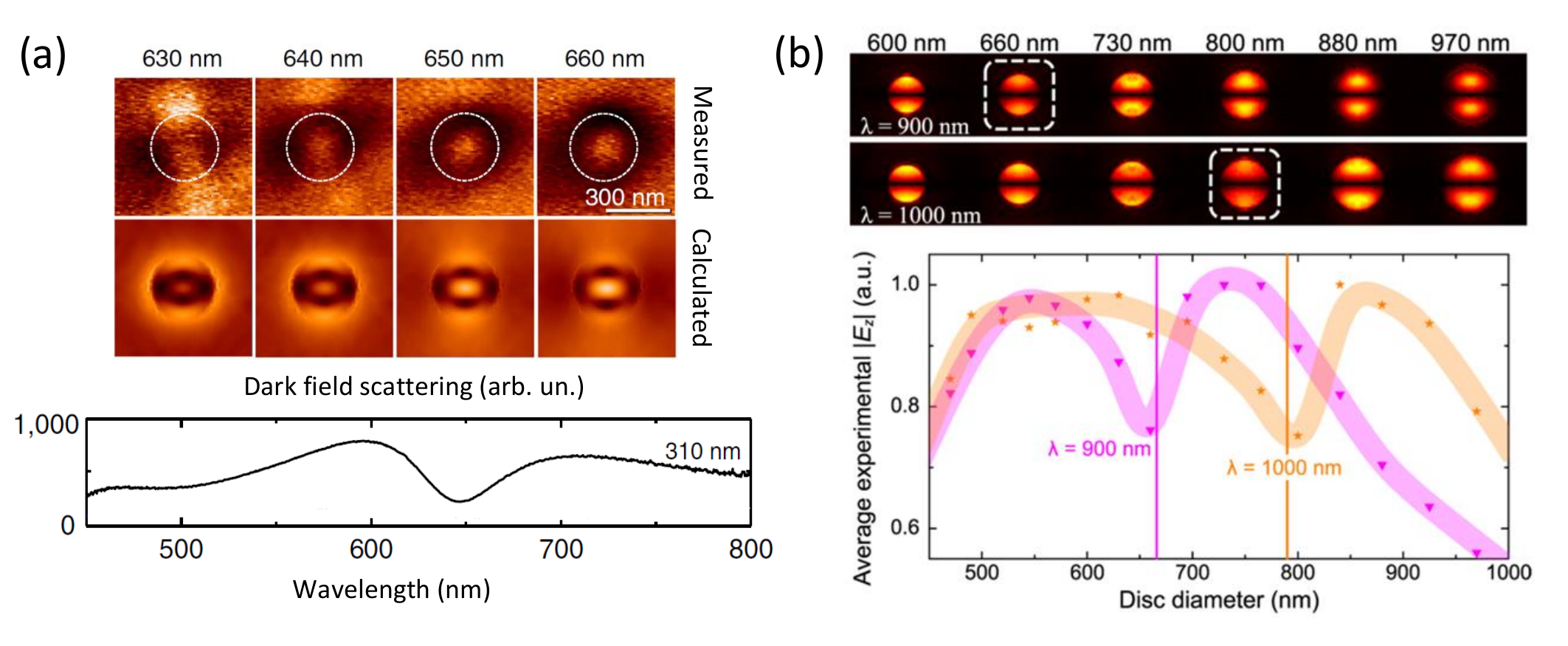}
 \caption{ {\bf Experimental observation of anapole states in all-dielectric nanoparticles}. (a) Experimental dark field scattering spectrum of a silicon nanodisk with a height of 50 nm and a diameter 310 nm. The top row shows experimental near-field scanning optical microscope (SNOM) measurements of the near-field enhancement around the silicon nanodisk close to the anapole wavelength of 640 nm. White dashed lines in the experimental images indicate the disk position.
(b) Near-field detection of anapole states. Experimental near-field $|E_z|$ distribution (top) and its average value for silicon disks, measured at two different illumination wavelengths: 900 (magenta triangles) and 1000 (orange stars) nm, disk diameters are labeled on top. The disks with the lowest near-field amplitude, encircled with the white dashed lines and marked by the vertical lines, support the second-order anapole states. Adapted from \cite{miroshnichenko2015nonradiating,zenin2017direct}.}
\label{fig:experiment}
\end{figure*}

First direct demonstration of the anapole state in the standalone dielectric nanoparticle was performed in 2015~\cite{miroshnichenko2015nonradiating}.
The first-order electric anapole, induced by the superposition of the electric and toroidal dipoles, was observed by SNOM measurements. By employing the multipole decomposition, it was numerically found that by changing the aspect ratio of a silicon nanodisk it is possible to detune all higher order multipoles away from the anapole excitation condition. 
Thus, for the optimized aspect ratio a silicon nanodisk exhibits a profound dip in the far-field scattering, which is accompanied by a near-field enhancement inside and around the disk [see Fig.~\ref{fig:experiment}(a)], 
associated with the anapole mode configuration.
 Experimentally, anapole was excited in a silicon nanodisk with a height of 50 nm and diameter 310 nm, that were fabricated on a quartz substrate using standard nanofabrication techniques. The disk becomes nearly invisible in the far-field at the wavelength of the anapole mode 640 nm. Both, near- and far-field measurements confirm the anapole mode excitation in the visible spectrum range [see Fig.~\ref{fig:experiment}(a)]. 
Here the anapole wavelength is far away from resonances and destructive interference between electric and toroidal dipoles happens in the local minimum of the fradoo
electric dipole moment.

Higher-order anapoles in the near infrared wavelength range were also later observed with silicon disks. The disks of 80 nm thickness and diameters varied from 470 to 970 nm were fabricated on a fused silica substrate~\cite{zenin2017direct}. 
In addition to the total scattering drops and local maxima in the accumulated electromagnetic energy, the anapole states were experimentally found to be characterized by a reduction of the normal near-field electric component and this property was successfully applied to detect the first two anapole states, as shown in Fig.~\ref{fig:experiment}(b)].  All three identification techniques provided similar results with deviations less than 5~\%. The higher-order anapole states were shown to possess stronger energy concentration and narrower resonances, which is advantageous for their applications.
 
Remarkable process of an active switching between dark (anapole) and bright (electric dipole) modes were demonstrated experimentally by the Bozhevolnyi {\it et al.} ~\cite{tian2018dynamic}. Ge$_2$Sb$_2$Te$_5$ (GST) with different degree of crystallinity allows tunability of the refractive index, and thus shifting positions of resonances supported by the nanoparticle in the broadband wavelength range. It was shown that increasing of a percent of crystalline phase in  a 450 nm radius-nanosphere from 0\% to 25\% leads to switching the electromagnetic response of the nanosphere from the electric dipole to the anapole state. Nanodisks made of GST were fabricated by using electron-beam lithography, magnet sputtering deposition, and standard lift-off process. Then higher-order anapole states in the nanodisks were experimentally identified,  suggesting much richer switching phenomena compared to the case of a symmetric GST sphere.
A broadband response of the switching effect was examined and it was shown that the ED-to-anapole transition in a GST nanodisk of arbitrary cristallinity can be achieved by simply introducing a phase change C = 50\% at any given wavelengths between 3.9 $\mu$m to 4.6 $\mu$m. 

\section{Applications of the anapole concept}

\subsection{Enhanced nonlinear effects} 

Engineered anapole states correspond to the cancellation of the far-field radiation in the first order and, consequently, to the  energy concentration in the subwavelength volumes, as was measured experimentally in Ref.~\cite{yang2018anapole}.  A tight field confinement associated with optical anapole modes can be utilized for enhancing nonlinear interactions at the nanoscale. 

The anapole state was first employed by Grinblat {\em et al.}~\cite{grinblat2016enhanced} for the observation of strong third-order nonlinear processes and efficient THG from Ge nanodisks. They observed a pronounced valley in the scattering cross section at the anapole mode, while the electric field energy
inside the disk is maximized due to high confinement within
the dielectric nanoparticle. Grinblat {\em et al.}~\cite{grinblat2016enhanced} investigated the
dependence of the third harmonic signal on the size of a
nanodisk and pump wavelength and revealed the main features of the anapole mode at a wavelength of 1650 nm, corresponding to an associated THG conversion efficiency of 0.0001, which is four orders of magnitude greater than the case of an unstructured germanium reference film. Furthermore, they concluded that the nonlinear conversion via the anapole mode outperforms that via the radiative dipolar resonances by about at least one order of magnitude, implying that the anapole-induced radiation suppression in this case plays a crucial role. 

\begin{figure*} [t!]
\centering \includegraphics[width=0.8\linewidth]{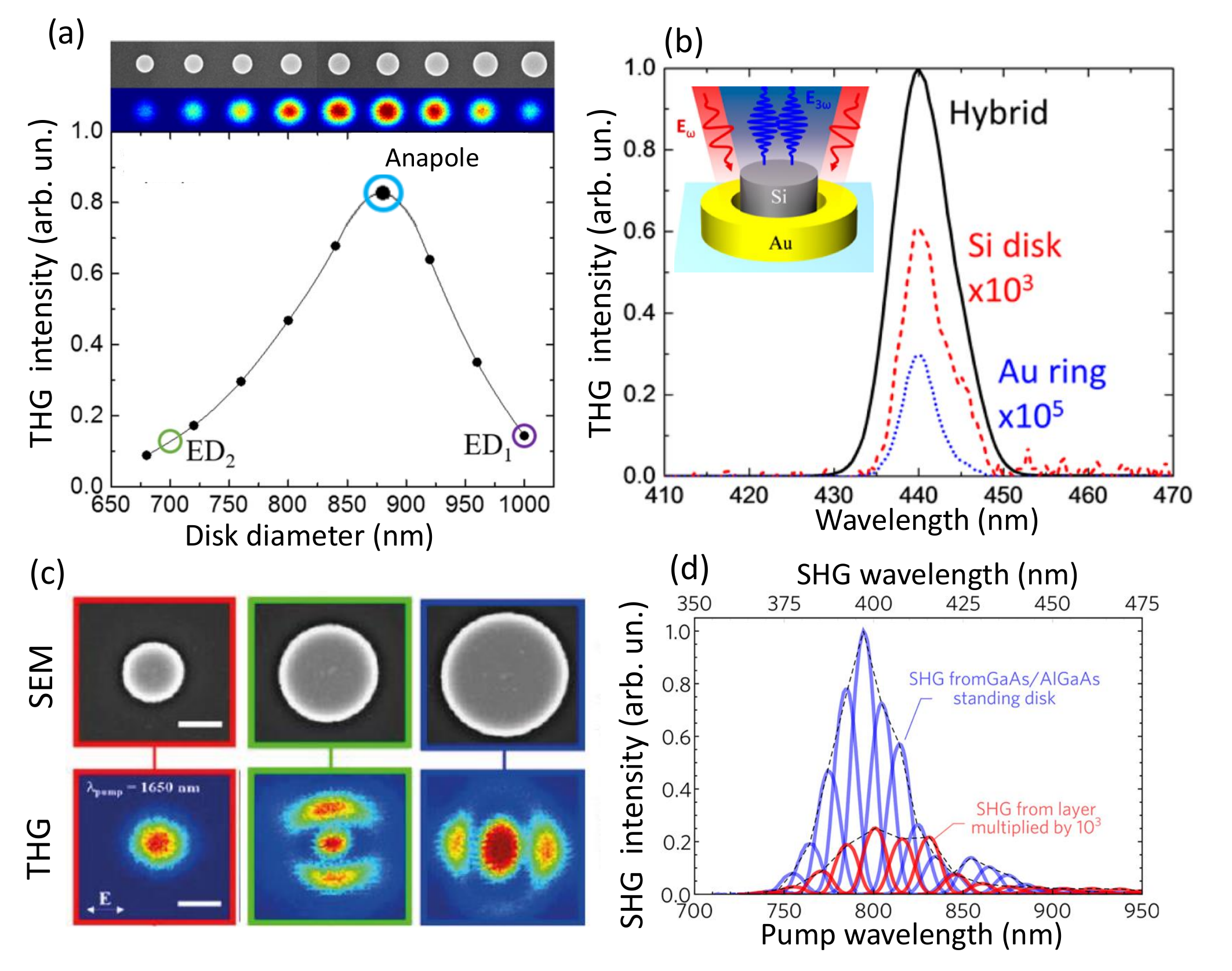} 
\caption{{\bf Nonlinear effects driven by 
the anapole states.} (a) Measured THG intensity with varying the disk diameter at the pump wavelength 1650 nm. The solid line connecting the dots is a guide to the eye. The corresponding SEM and THG intensity maps are displayed on the top. (b) THG spectra of the hybrid metal-dielectric structure (shown in the inset), isolated Si nanodisk, and bare Au nanoring measured at 1320 nm excitation wavelength and 1 $\mu$W excitation power. (c) SEM micrographs (top) and THG images (bottom) of Ge nanodisks with radii 350 nm (left), 550 nm (middle) and 700 nm (right). from left to right: the anapole state, the higher-order anapole, the non-anapole state. White scale bars marks 500 nm length. (d) Wavelength dependence of the SHG spectra from the GaAs/AlGaAs disk (blue lines) and epitaxial layer (red lines, signal is multiplied by 10$^3$), the dashed gray lines are the SHG spectra envelopes; recorded by sweeping the excitation laser wavelength from 690 nm to 900 nm with 10 nm step. Adapted from \cite{grinblat2016enhanced,shibanuma2017efficient,grinblat2016efficient,timofeeva2018anapoles}.}
\label{fig:THG}
\end{figure*}

The THG conversion efficiency can be further enhanced by the geometry optimization~\cite{shibanuma2017efficient,gili2018metal}. For example, in a metal-dielectric nanostructure composed of a Si nanodisk as a core and a gold ring as a shell, THG conversion efficiency via the anapole state reaches 0.007\% at the third-harmonic wavelength of 440 nm [see Fig.~\ref{fig:THG}(b)]. Specifically, in this geometry a Si disk plays the role of the nonradiating particle with the anapole-induced enhancement of the field, and the gold ring boosts the electric field enhancement inside the Si nanodisk due to its plasmonic properties. Similarly, in the work by the Celebrano {\it {et al.}} ~\cite{gili2018metal} hybrid metal-dielectric structures were studied experimentally as efficient frequency converters. AlGaAs nanopillar surrounded by a gold ring support the anapole mode, consisted of equal electric dipole response of the ring and toroidal response of the pillar. This enables the improvement of both the second- and third-order nonlinear efficiencies, with measured enhancement factors of about 30 and 15 for the second and third harmonics generation processes, respectively.

Grinblat {\em et al.}~\cite{grinblat2016efficient} studied the effect of the high-order anapoles on the THG intensity, and compared the anapoles of different orders. They used arrays of Ge nanodisks with different radii, and varied the radius of each nanodisk from 300 nm to 700 nm keeping the fixed disk height equal to 200 nm. This allowed scanning the modes excited in the nanodisk and alongside changing the distribution of the electromagnetic fields inside the nanoparticles. In the nanodisk with the radius 350 nm at the wavelength 1650 nm, Grinbrat {\em et al.}  observed the first-order (fundamental) anapole state with a typical distribution of electromagnetic fields similar to those in Fig.~\ref{fig:THG}(c) and discussed above.
In the nanodisk of radius 550 nm, they observed the generation of a second-order anapole state discussed previously in Ref.~\cite{zenin2017direct}. THG efficiencies from the anapole states of the first- and second- orders were compared to each other and to that from a 700 nm-radius nanodisk. The latter has quite similar distribution of electromagnetic fields inside the nanoparticle [see Fig.~\ref{fig:THG}(c)] but it corresponds not to a minimum but to a maximum of the overall extinction. THG from all those modes was measured at the pump wavelength 1650 nm. It was established experimentally that the third-harmonic intensity is higher by orders of magnitude in the case of the second-order anapole state, being in agreement with results of Ref.~\cite{zenin2017direct} and suggesting that the energy concentration increases as the anapole state order grows. As a result, high THG conversion efficiency of up to 0.001\% to the wavelength of 550 nm was achieved.

Efficiency of THG can be also significantly increased by placing a dielectric anapole resonator on a metallic mirror. Xu {\em et al.}~\cite{Xu2018Boosting} demonstrated experimentally that this can enhance the third-harmonic radiation intensity by additional two orders of magnitude compared to a typical anapole resonator located on an insulator substrate.

\begin{figure*} [t]
\centering
\includegraphics[width=0.9\linewidth]{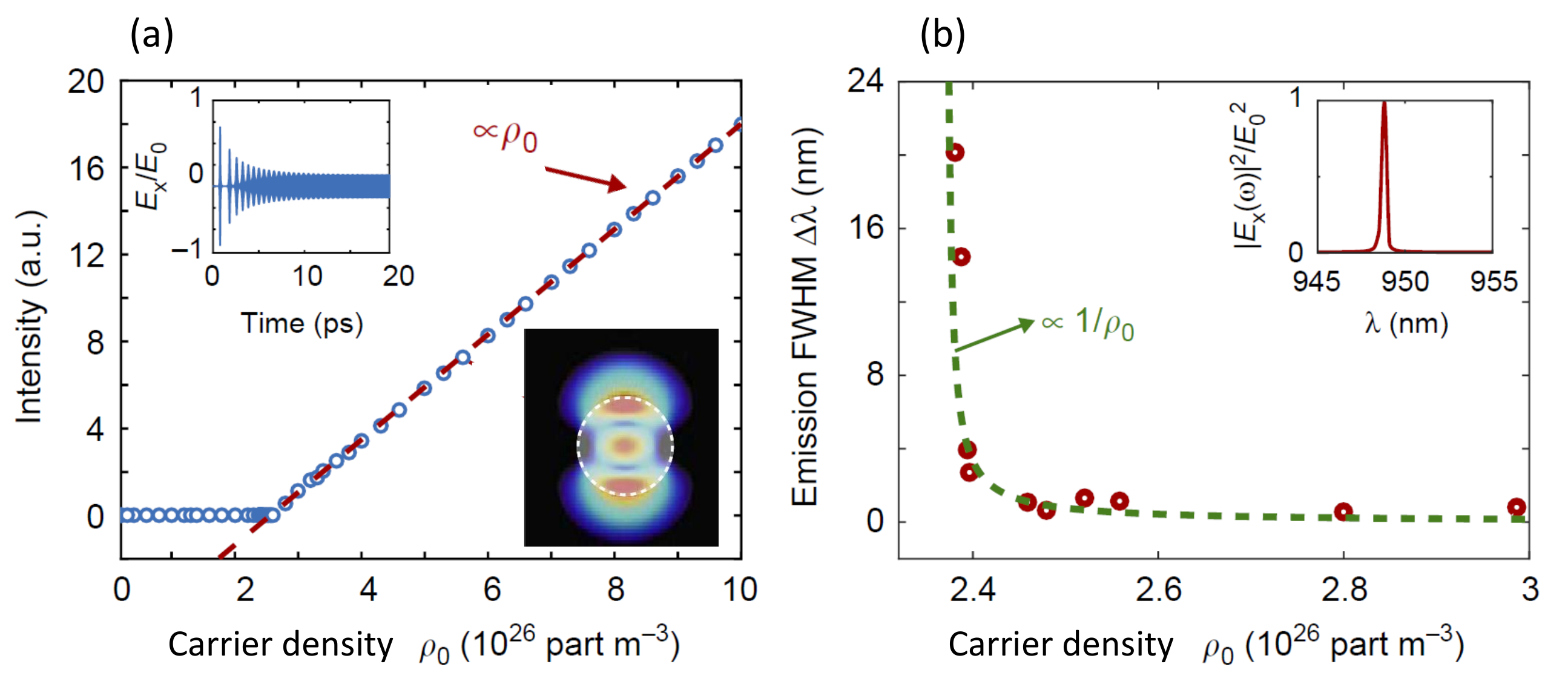}
\caption{{\bf Operation of the anapole nanolaser}. (a) Input/output diagram of anapole amplitude vs. carrier density $\rho_0$. Upper inset: Time trace of the field component $E_x$. Lower insert: Cross-section of the electromagnetic energy in the steady-state. (b) Corresponding linewidth behavior and (inset) power spectral density measured inside the nanodisk. Adapted from \cite{gongora2017anapole}.}
  \label{fig:laser}
\end{figure*}

While silicon is a centrosymmetric material and possesses strong cubic nonlinearity, its quadratic nonlinearity is inhibited in the bulk and second-harmonic generation (SHG) can be generated mostly from the interfaces. For efficient second-order nonlinear applications, nanostructures made of high-permittivity noncentrosymmetric dielectric materials, such as A$_{III}$B$_{V}$ semiconductors, can be advantageous. In Ref.~\cite{timofeeva2018anapoles} 
SHG intensity at the anapole state in the geometry of vertically standing GaAs/AlGaAs pillars on a substrate was demonstrated to exceed that in the bare layer up to 5 times [see Fig.~\ref{fig:THG}(d)].
Rocco {\em et al.}~\cite{rocco2018} studied the effective coupling of electric and toroidal modes in AlGaAs nanodimers to enhance the SHG efficiency with respect to the case of a single isolated nanodisk. They have demonstrated that proper near-field coupling can provide further degrees of freedom to control the polarization state and the radiation diagram of the second-harmonic fields. 

Similarly, the anapole state can enhance other nonlinear processes, in particular, the Raman response of Si nanoparticles, as shown recently both theoretically and experimentally~\cite{baranov2018anapole}. The Stokes and anti-Stokes Raman spectra from Si nanoparticle arrays with varying disk radius were collected using excitation at 785 nm wavelength and revealed a pronounced phonon peak at 522 cm$^{-1}$. The largest enhancement of the Stokes emission, the factor of $\approx$80 compared to an unstructured Si film, was observed for the radius of 190 nm. Nanodisks with this radius support the anapole state at the excitation wavelength of 785 nm, manifested as minima in extinction and reflectivity.

\subsection{Nanolasers in subwavelength photonics} 

The concept of coherent light emitters with the nanoscale sizes has attracted a lot of attention especially during last decade due to the recent technological progress in lab-on-chip fabrication technologies. Nanolasers are usually associated with plasmonics where the subwavelength localization of light is driven by the physics of surface plasmon polaritons. Such spaser-type nanolasers have been fabricated and studied experimentally~\cite{noginov2009demonstration,lu2012plasmonic}. Dielectric photonic structures have advantages due to low ohmic losses of dielectric materials, but they usually require much larger scales such as those provided by semiconductor nanowires~\cite{huang2001room,chen2011nanolasers}.
Application of anapoles to the physics of nanolasers looks especially attractive because such states do not dissipate energy in the far-field region and effectively accumulate energy inside the nanoparticle, which can increase the local state density and subsequently enhance the overall laser efficiency. By harnessing the non-radiating nature of the anapole state, one can engineer nanolasers as on-chip sources with unique optical properties. 

Recently suggested {\em an anapole nanolaser}~\cite{gongora2017anapole} is based on a tightly confined anapole mode produced by interference of the dipole mode and toroidal mode in an optically pumped semiconductor nanodisk (see also Refs.~\cite{gongora2016near,gongora2017engineering}).
Importantly, anapole modes can be employed to
resolve a major challenge with the efficient coupling of light
to nanoscale optical structures.
Indeed, in the paper~\cite{gongora2017anapole}, Gongora {\em et al.} studied an advanced theoretical model of spontaneously polarized nanolaser able to couple light into waveguide channels with four orders of magnitude intensity higher than that for classical nanolasers. Moreover, the anapole nanolaser allows generating ultrafast (of 100~fs) pulses via spontaneous mode locking of several anapole modes. The anapole laser does not require extended structures and the radiationless state is excited with no additional radiating components. A resonator employing the anapole mode is based on InGaAs semiconductor. 

Figures~\ref{fig:laser}(a,b) show the intensity and linewidth of the electromagnetic field of the anapole amplified inside the nanodisk of In$_{0.15}$Ga$_{0.85}$As with diameter 440 nm and height 100 nm. The behavior of the anapole amplitude versus pumping rate [see Fig. \ref{fig:laser}(a,b)] is that of a standard laser, with a linear relationship between pumping rate and amplified intensity [see Fig. \ref{fig:laser}(b)] he anapole laser also shows an equivalent Schawlow–Townes linewidth [see Fig. \ref{fig:laser}(b] that quickly reaches a stable value of $\approx$2nm. A characteristic time evolution of the electric field is displayed in Fig.~\ref{fig:laser}(a) (see the upper inset) that shows the reaching of a stable stationary emission state after an initial transient. The corresponding intensity spectrum in Fig.~\ref{fig:laser}(b) further confirms that the emission corresponds to the amplification of the anapole state at 948nm. Based on this nanolasers it is possible to construct on-chip ultrafast pulse generation due to anapole mode locking, as described in Ref.~\cite{gongora2017anapole}. 

Coupled anapoles can provide a new path  for nonradiative energy transfer~\cite{liu2017high,Mazzone2017}. 
Due to exotic near-fields, coupling and hybridization of anapoles obey different rules compared to electric and magnetic dipoles. %
In nanodisk arrays the $Q$-factor can me made about one order larger than that of the single disks associated with the nonradiating anapole modes and the collective oscillations of the arrays~\cite{liu2017high}. The coupling between the two interfering modes can be modified and  $Q$-factor further enlarged by introducing split in the disks and adjusting its gap width.
When the resonance energies of the electric dipole mode and the 
The Ref.~\cite{Mazzone2017} describes sub-wavelength guiding via near-field transfer of anapole states. Due to the near-field confinement produced by the anapole state, the anapole nanochain appears robust against bending and splitting of the integrated waveguide. This opens a way to the realization of integrated splitters and 90-degree bends without parasitic radiation losses.

\subsection{Anapole-driven metamaterials} 

\begin{figure*}[t]
	\centering
	\includegraphics[width=0.9\linewidth]{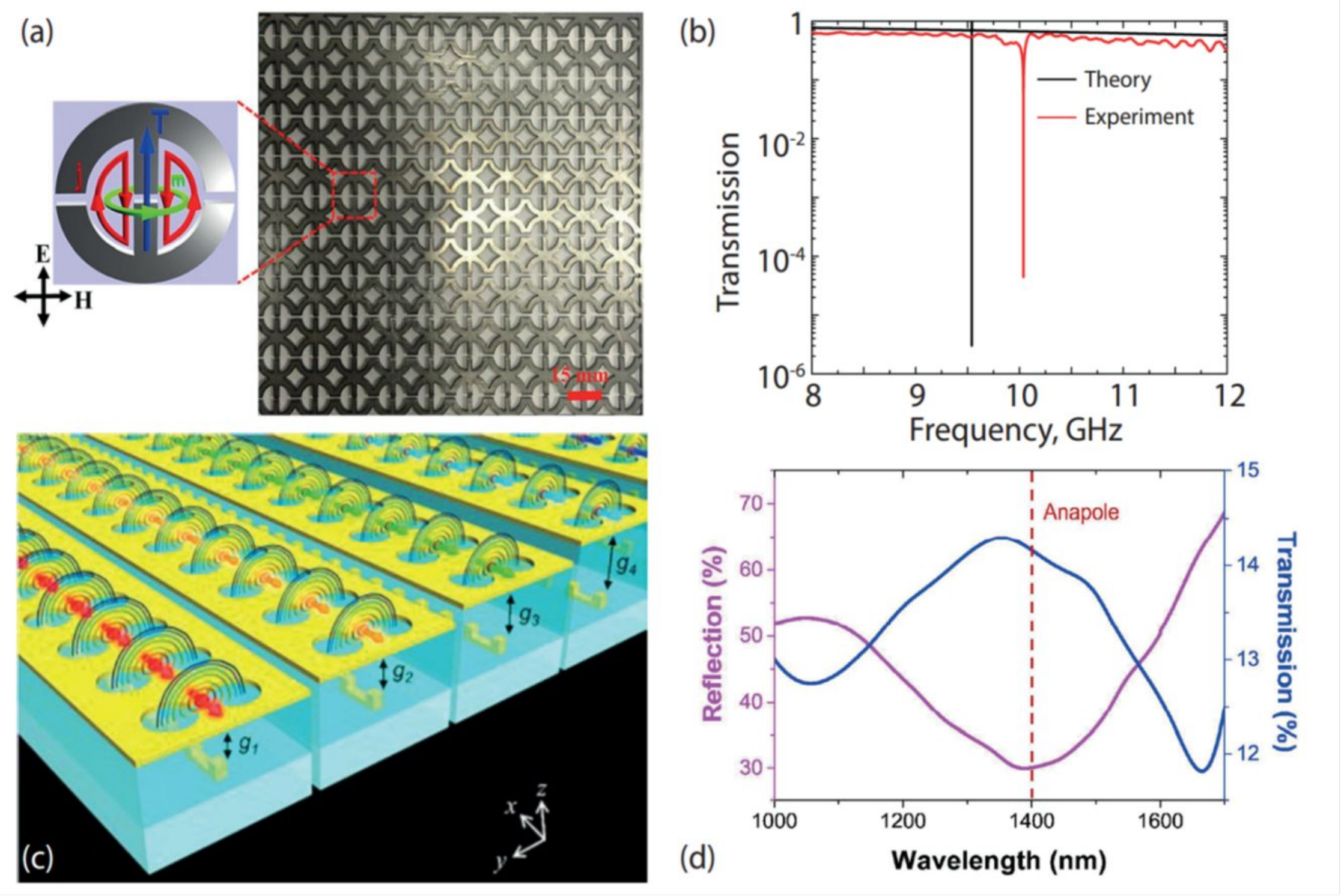}
  \caption{{\bf Designs of anapole metamaterials.} (a) Left: a unit cell of a metamaterial supporting toroidal dipolar excitation. Red arrows: displacement currents ${\bf j}$ induced by the vertically polarized plane wave, blue arrow: induced toroidal dipole moment ${\bf T}$ of the metamolecule, green round arrow: circulating magnetic moment ${\bf m}$. Right: the metamaterial sample. (b) Transmission spectra, calculated (black lines) and experimental (red lines), for the sample shown in Fig.~\ref{fig:metamaterial}(a). (c) Schematic of the anapole metamaterial, consisting of a gold film perforated by dumbbell-shaped apertures, situated above a planar array of VSRRs suspended in a dielectric spacer layer. The color arrows and the streamlines show the orientation of induced toroidal dipoles and the corresponding magnetic field distribution, respectively. (d) Measured far-field spectra of the plasmonic anapole metamaterial illustrated in Fig.~\ref{fig:metamaterial}(c) under $x$-polarized illumination. The dark-red dashed line marks the resonant wavelength of anapole mode obtained numerically. Adapted from \cite{basharin2017extremely,wu2018optical}.}
  \label{fig:metamaterial}
\end{figure*}

The anapole excitation has been firstly experimentally
realized within engineered composite metallic metamaterials in the microwave regime, where the
excitation of other multipoles is negligible~\cite{Fedotov2013}.
In Ref.~\cite{basharin2017extremely} the electromagnetic response of the metamaterial consisting of planar conductive metamolecules formed by two symmetric split rings (see Fig.\ref{fig:metamaterial}(a)) was shown to exhibit an extremely narrow peak in transmission spectrum (Fig.\ref{fig:metamaterial}(b) at the frequency where toroidal moment interferes destructively with suppression of electric dipole radiation.
In the paper~\cite{wu2018optical} the anapole metamaterial operating in the IR range was designed based on plasmonic nanostructures, Fig.\ref{fig:metamaterial} (c). Metaatom composed of a dumbbell aperture in a gold film and a vertical split-ring resonator exhibits close to ideal suppression of electric dipole radiation. The remarkable advantage of this design is that in this structure the electric field concentrates in the imaginary torus, bounded by the dumbbell aperture, so that the area of electric field localization is accessible, which is highly relevant to sensing applications, such as the surface-enhanced Raman scattering (SERS).  The efficient coupling between anapole meta-materials and molecules supporting toroidal resonances in natural substances suggests feasibility of toroidal spectroscopy. Depending on the excitation wavelength, the toroidal dipole or anapole state can be clearly identified in response spectra. At the anapole wavelength, suppression of reflection coefficient down to 30 $\%$ with an increase of transmission coefficient up to 14.5 $\%$ is observed experimentally, Fig.\ref{fig:metamaterial} (d).

Anapoles in particles and metamolecules made of dielectric materials opens a route to low-loss metamaterials and metadevices.
The feasible design  of all-dielectric anapole metamaterial in the optical range was suggested theoretically in the paper~\cite{basharin2018anapole}. The proposed structure consists of arranged metamolecules - clusters of four holes perforated in a silicon slab and can be fabricated by Focused Ion Beam techniques. The anapole state in this system corresponds to the peak up to 100 \% in transmission spectrum and its position depend on the parameters of the structure. Both gradual increasing an angle of illumination and reducing the depth of holes result in a redshift of transmission peak, whereas a resonance width decreases as the angle of incidence grows and, in opposite, increases with the hole deepening. 

\vspace{-0.5 cm}
\subsection{Other applications of optical anapoles} 

If scattering by some object is mainly determined by its electric dipole component, the anapole concept can be applied for cloaking of this object. For instance, if a toroidal response of the shell can fully compensate electric dipole response of camouflaged object in the core, the overall system becomes invisible. This approach was discussed in the paper~\cite{ospanova2018multipolar} theoretically with an example of a system of cylinders. 
The electric dipole response of a small cylinder, placed in the center of the system, can be camouflaged by the toroidal response of the shell, a cluster of four other cylinders exhibiting a strong toroidal response.

Anapole electromagnetic state can be treated as an ideal suppression of the selected multipolar order but not as a completely non-radiating source. If radiation of some multipoles is strongly suppressed, other multipoles can still radiate efficiently. In fact, at the first-order anapole, the pure magnetic dipole source can be realized, as theoretically suggested in Ref.~\cite{feng2017ideal}. Scattering pattern of an Au core -- Si shell nanodisk embedded in air with  optimized dimensions was shown to approach the ideal magnetic dipole radiation pattern. 
Multipole decomposition proved the contributions of the electric dipole, electric quadrupole and magnetic quadrupole to the total scattering to be negligible, while slight deviations from the ideal magnetic dipole caused by weak higher-order multipoles.

Anapole states can also be used for the formation of the electric and magnetic near-field hotspots in nanostructures. The electromagnetic energy is transformed from the far field to the tightly localized fields, thus being strongly enhanced. In Ref.~\cite{yang2018anapole} it is demonstrated that individual all-dielectric nanostructure can trap the fields with intensity enhancement exceeding 3 orders of magnitude. This remarkable effect is achieved with a Si nanodisk supporting an anapole state by introducing a high-contrast slot area.  Possibility to enhance pure magnetic hotspots with effect of spatial separation of electric and magnetic fields based on the electric dipole suppression and electric dipole enhancement was described in Refs.~\cite{baryshnikova2017optimization,terekhov2016nonradiating,Cojocari2018spatial}. 

The efficient excitation of nonradiating anapoles can play a significant role for  enhancing light absorption in photonic nanostructures, e.g. nanodisks~\cite{Wang2016absorption} and core-shell particles~\cite{Wang2017absorption}. 
In Ref.~\cite{Wang2016absorption} the authors theoretically demonstrated wide and controllable wavelength tunability of anapole-driven absorption enhancement in dielectric nanodisks and square nano-pixels, and optimum conditions to achieve the highest absorption rate on the particle sizes and excitation 
were identified.  Thus, while for standalone germanium nanodisk absorption rate is higher than 5 in a bandwidth about 200 nm, for heterostructure combined from two different nanodisks this bandwidth is about 800 nm. The nanostructured arrays of particles were further designed to increase the absorption bandwidth that can serve as functional units in  semiconductor photodetectors. 

A promising application of the anapole concept can be found in the study of interaction of an anapole with molecular excitons and other molecular excitations. In the paper~\cite{Liu2018exciton} a nanostructure consisting of silicon disk as a core and aggregate of molecules as a coaxial shell was examined. Molecules in the aggregate possess exciton excitations. The resonant coupling is evidenced by a scattering peak around the exciton transition frequency, and the anapole mode splits into a pair of eigenmodes characterized by pronounced scattering dips. These observed dips are associated with the formation of two hybrid anapole modes caused by the coherent energy exchange in the heterostructure. Molecules located around the apexes of the disk perpendicular to the incident polarization play the dominant role in this interaction, and a special near-field distribution characteristic to the anapole state allows increasing an upper limit for the width of the aggregate ring to enhance the resonance coupling. %

\section{Related interference effects}

The concept of the anapole states represents a special, yet the most intriguing, example of interference effects for isolated subwavelength dielectric particles. During recent years, several related concepts have been introduced and developed. The associated multipolar effects show their importance for numerous applications of all-dielectric Mie-resonant photonics.

The study of multipolar interference goes back to Kerker~\cite{kerker} who considered interference between electric and magnetic dipoles for a hypothetical magnetic sphere. Rejuvenated by the recent explosive development of the field of metamaterials and especially its core concept of optically-induced artificial magnetism, the Kerker effect has gained an unprecedented impetus and rapidly pervaded different branches of nanophotonics. At the same time, the concept behind the effect itself has also been significantly expanded and generalized, as was described in the recent review paper~\cite{liu_oe} not only for the scattering by individual particles and their clusters but also for the manipulation of reflection, transmission, diffraction, and absorption with metagratings and metasurfaces.

\begin{figure}[t!] 
\centering\includegraphics[width=0.99\linewidth]{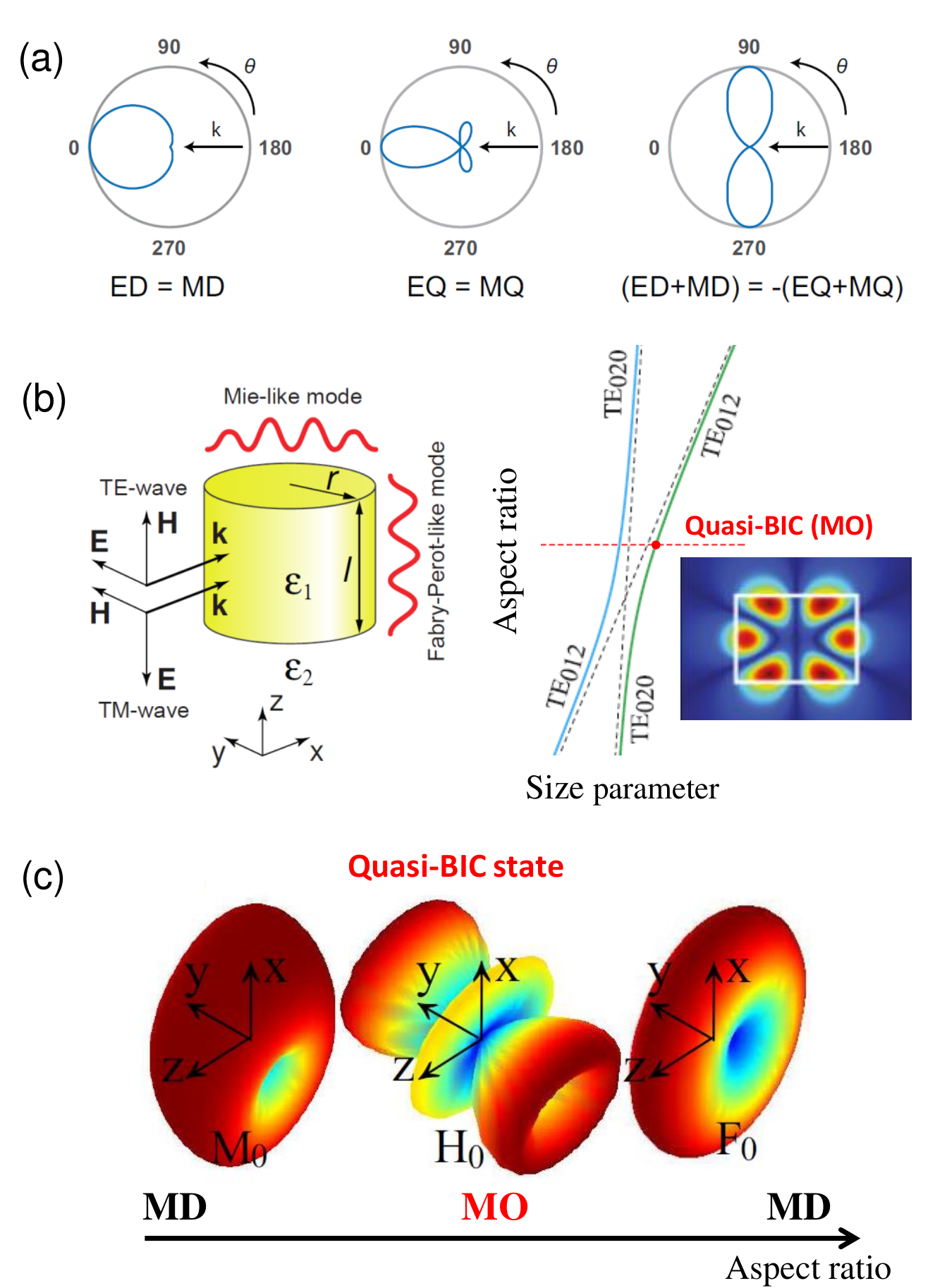} 
 \caption{  {\bf Examples of novel interference effects in isolated nanoparticles.} (a)  Concept of the formation of an ideal transverse scattering pattern. Electric dipole (ED) is in-phase  with a magnetic dipole (MD) and an electric quadrupole (EQ) is in-phase with a magnetic quadrupole (MQ), while the dipoles are out-of-phase with the quadrupoles~\cite{shalin}. (b) Strong coupling of modes in a subwavelength dielectric resonator with an emergence of quasi-BIC state~\cite{bic_prl}. (c) Example of radiation pattern transformation for the mode when passing the anti-crossing point~\cite{bic_wei}. The quasi-BIC supercavity mode condition corresponds to the magnetic octupole radiation. Adapted from \cite{shalin,bic_prl,bic_wei}. 
}
\label{fig:related}
\end{figure}

Interplay between electric and magnetic dipolar resonances makes it possible to achieve either constructive or destructive interference leading to remarkable scattering properties of subwavelength dielectric particles in the forward or backward directions~\cite{kerker, liu_oe}. Overlapping electric and magnetic multipoles of higher orders further suggests advanced strategies for shaping light radiation. Recently Shamkhi {\it et al.} ~\cite{shalin} described the transverse scattering of light by high-index subwavelength particles with the simultaneous suppression of both forward and backward scattering. This unusual effect occurs when the in-phase electric and magnetic dipoles become out of phase with the corresponding quadrupoles, as shown schematically in Fig.~\ref{fig:related}(a). By applying the Mie theory, they have obtained the underlying conditions, and  provided experimental data for the microwave spectral range.

Another special type of interference originates from strong coupling between the leaky modes supported by a single subwavelength high-index dielectric resonator. Rybin {\it et al.}~\cite{bic_prl} demonstrated that the strong coupling regime results in resonances with high-quality factors ($Q$ factors), which are related to the physics of bound states in the continuum (BIC) when the radiative losses are almost suppressed due to the Friedrich–Wintgen scenario of destructive interference. One type of interfering modes is formed mainly due to reflection from a side wall of the cylinder, and they can be traced to the Mie resonances of an infinite cylinder, see Fig. 6(b). The other type of modes is linked to the Fabry-Perot-like resonances, so that, for a cylindrical resonator, the strong coupling between the Mie-like and Fabry-Perot-like modes is clearly manifested as avoided resonance crossing points with the appearance of the quasi-BIC states. Additionally, Rybin {\it et al.}~\cite{bic_prl} revealed a link between the formation of the high-$Q$ resonances and peculiarities of the Fano parameter behavior in the scattering cross-section spectra.

The formation of quasi-BICs can be naturally understood through multipolar transformations of coupled modes~\cite{Poddubny2018quantumBIC,bic_wei}. For a nonspherical shape of the resonator, the interacting modes can be in general viewed as superpositions of multipoles.
For the case of Figs.~\ref{fig:related}(b,c), a basis of parent multipoles is constituted mainly by magnetic dipoles and octupole. 

The occurrence of high-$Q$ supercavity mode is accompanied by increasing the order of a dominating multipole from $l=1$ (MD) to $l=3$ (MO) and corresponds to the decoupled magnetic octupole. At quasi-BIC condition, two magnetic dipole modes interfere destructively in the coupling to the octupolar mode, thus, restoring its high quality factor~\cite{Poddubny2018quantumBIC}. 

As shown in Ref.~\cite{Poddubny2018quantumBIC} in the framework of the three-state model, such mutual interplay of parent multipoles essentially determines generation of both high-$Q$ resonant states and dark states in dielectric nanoresonators. Relying on this instructive interpretation and generalized Kerker effects of interferences among different electromagnetic multipoles, the radiation of the subwavelength high-$Q$ supermodes can be made unidirectional by breaking symmetries of the dielectric scatterers~\cite{bic_wei}, that  can be useful for nanoscale lasing and sensing related applications. Similar to the anapole states, such BIC-inspired modes can be also employed for a substantial enhancement of nonlinear and quantum effects at the nanoscale~\cite{bic_nonlinear,Poddubny2018quantumBIC}.

\section{Concluding remarks}

We have presented a brief overview of the theoretical and experimental results on the rich physics of optical anapoles,  putting an emphasis on the recent progress in its application to subwavelength dielectric structures and meta-optics. Being a universal physical model of non-scattering objects and non-radiating sources, the anapole mode can be accessed in nanophotonics by using conventional dielectric materials. Dielectric nanoparticles can support almost radiationless anapole modes allowing an interference of the toroidal and electric or magnetic multipoles through a geometry tuning, with a pronounced dip in the far-field scattering accompanied by the specific near-field distribution. The anapole properties, particularly, tightly confined fields, have important implications for modern nanophotonics beyond what is expected from the dipole approximation.

The anapole concept, as well as related interference effects in nanoparticles, offers an attractive platform for an efficient control of light-matter interaction in various nanoscale functional photonic elements and devices, including broadband absorbers, efficient nonlinear light sources, nanolasers, and ultrasensitive biosensors. Novel horizons in applications driven by the controllable anapole response, such as non-invasive sensing, broadband photodetectors, near-field lasing, spurious scattering suppression and optical invisibility, can be explored in the near future in all-dielectric resonant photonics. 

\section*{Funding Information}

This work was supported by the Goverment of the Russian Federation (Grant 08-08), the Russian Science Foundation (project 17-72-10230), the Russian Foundation for Basic Research (Grant 18-02-00381), the Australian Reserach Council, and the Strategic Fund of the Australian National University. 

\section*{Acknowledgments}

The authors acknowledge useful suggestions and comments from A. Basharin, A. Evlyukhin, A. Fratalocchi, and Wei Liu, and they are also thankful to all colleagues and co-authors of joint papers for valuable contributions.




\end{document}